\documentclass[twocolumn,showpacs,preprintnumbers,amsmath,amssymb,prb]{revtex4}

\usepackage{graphicx}%
\usepackage{dcolumn}
\usepackage{amsmath}

\makeatletter
\def\btt#1{\texttt{\@backslashchar#1}}%
\DeclareRobustCommand\bblash{\btt{\@backslashchar}}%
\makeatother

\begin{document}

\title{
The effect of uniaxial pressure on the magnetic anomalies of the\\
heavy-fermion metamagnet CeRu$_2$Si$_2$\\
}

\author{
S. R. Saha, H. Sugawara, T. Namiki, Y. Aoki, and H. Sato\\
}

\affiliation{
Department of Physics, Graduate~School~of~Science,~Tokyo~Metropolitan~University,
Minami-Ohsawa 1-1,~Hachioji,~Tokyo~192-0397,~Japan
}

\date{\today}

\begin{abstract}
~~~The effect of uniaxial pressure ($P_u$) on the magnetic susceptibility ($\chi$),
magnetization ($M$), and magnetoresistance (MR) of the heavy-fermion metamagnet
CeRu$_2$Si$_2$ has been investigated. For the magnetic field along the tetragonal
$c$ axis, it is found that characteristic physical quantities, i.e., the temperature of
the susceptibility maximum ($T_{\rm max}$), the pagamagnetic Weiss temperature
($\Theta_p$), $1/\chi$ at 2 K, and the magnetic field of the metamagnetic anomaly
($H_M$), scale approximately linearly with $P_u$, indicating that all
the quantities are related to the same energy scale, probably of the
Kondo temperature. The increase (decrease) of the quantities for
$P_u\parallel c$ axis ($P_u\parallel a$ axis) can be attributed
to a decrease (increase) in the nearest
Ce-Ru distance. Consistently in MR and $\chi$, we observed a sign that the anisotropic
nature of the hybridization, which is believed to play an important role in the
metamagnetic anomaly, can be controlled by applying the uniaxial pressure.\\
\end{abstract}
\maketitle
%
\section{Introduction}
~~~The compound, CeRu$_{2}$Si$_{2}$, crystallizing in the tetragonal
ThCr$_{2}$Si$_{2}$-type structure, is one of the most intensively
studied and best characterized systems
among the heavy-fermion compounds.\cite{haen} The reason is that apart from a large
electronic specific-heat coefficient $\gamma$ $\sim$ 360 mJ/mol K$^{2}$ in zero
field,\cite{fisher,aoki1} there is an abrupt nonlinear increase
of magnetization, the so called metamagnetic anomaly (MA)
from a paramagnetic ground state,
around an external magnetic field of $H_M\sim$ 80 kOe (Ref. 1)
applied only along the $c$ axis of the tetragonal structure below 15 K.
Not only the magnetization process but also many other physical properties
have been reported to be anomalous in this field region around
$H_M$.\cite{aoki1,haoki1,sakakibara1,kambe12}\\
~~~The origin of the anomaly is still a matter of debate, despite intensive
investigations. The de Haas-van Alphen effect studies have shown that
both the Fermi surface and the
effective mass change considerably around $H_M$, suggesting a change of
4$f$-electron character from itinerance in the low-field state to localization
in the high-field state.\cite{haoki1}
In contrast, low-temperature magnetization measurements\cite{sakakibara1}
suggest that the low-field state is continuously connected to the high-field state
across $H_M$. Moreover, a peak observed in the Hall resistivity at $H=H_M$
disappears on approaching $T=$ 0, suggesting no abrupt change in the
Fermi surface.\cite{kambe12}
Reports of hydrostatic pressure experiments\cite{voiron1,mignot12} reveal that the
volume reduction enhances the characteristic energy of the quasiparticle system
and leads to a drastic shift of $H_M$ to higher fields. From this large effect
of pressure a very large electronic Gr\"uneisen
parameter $\sim$ 185 Mbar$^{-1}$ has been inferred.
The anisotropic hybridization between 4$f$ and conduction electrons
leading to an anomalous peak in the quasiparticle density-of-states (DOS)
is argued to play an important role in the MA; the MA appears when the peak crosses
the Fermi level at high magnetic fields.\cite{aoki1,hanzawa1ohara1}
However, no direct evidence of the anisotropic hybridization effect can be
obtained from hydrostatic pressure experiments. The uniaxial-pressure experiment
has the potential for providing useful information on anisotropic hybridization.
In view of these reasons, we have investigated the effect of uniaxial pressure
in CeRu$_2$Si$_2$ with magnetic and transport experiments.\\
%
\section{Experiment}
~~~Single crystals of CeRu$_2$Si$_2$ were grown by the Czochralski pulling method
in a tetra-arc furnace with an argon atmosphere. The single crystalline nature was
confirmed by back-reflection-Laue techniques. The high quality of the
single crystal was inferred from the residual resistivity ratio $\geq$ 110.
Electrical resistivity and magnetoresistance were measured by the standard dc four
probe method using a computer-controlled current source and nanovoltmeter (182 Keithley),
using a top-loading $^3$He cryostat equipped with a 160 kOe superconducting magnet
(Oxford Instruments Co., Ltd.). The electrical contacts of Ag current leads were
affixed to the sample by indium soldering.  Au wires of 80-$\mu {\rm m} \phi$
were spotwelded to the sample as voltage leads. Uniaxial pressures were
generated by using a piston-cylinder-type CuBe pressure cell for transport experiments,
recently designed and constructed by us\cite{sahaupcell}, so as to suit
the above cryostat. Single-crystal samples
(typical dimensions: $\sim$ 0.7 x 1 x 2 mm$^3$) were sandwiched between
two disc-shaped ZrO$_2$ plates, which provided the
electrical isolation. A Nitflon tape of 0.08-mm thickness was placed
between the sample and
ZrO$_2$ plates in order to prevent the breaking of the sample due to its surface
roughness under pressure, if any. Uniaxial pressures were applied on a ZrO$_2$ ball,
placed inside a ring-shaped guide on the ZrO$_2$ plate above the sample, which prevented
any rotation of the sample under pressure. Uniaxial pressures produced on the sample at
low temperatures were calibrated by measuring the superconducting transition
temperature of Sn placed in the cell by an induction method.
Magnetic properties were measured by a commercial superconducting quantum interference
device (SQUID) magnetometer. Uniaxial pressures parallel to the magnetic fields were
generated by using a modified version of the SQUID-pressure cell reported by
Uwatoko {\it et al.}\cite{uwatoko1}
Uniaxial pressures produced on the rectangular shaped single crystal
($\sim$ 1.5 x 1.5 x 2 mm$^3$) at low
temperatures were calibrated by measuring the  Meissner effect  of a small
piece of Pb, placed in the pressure cell. The known pressure dependences of the
superconducting transition temperature of Sn (Ref. 12)
(for transport measurements) and Pb (Ref. 13) were used for these purposes.
Uniaxial pressures perpendicular to the magnetic fields were generated by a
uniaxial-pressure cell recently designed and constructed by us\cite{sahaupcell}.
In this case, the uniaxial pressures were determined at room temperature
by the absolute value of force applied on the sample. The total magnetization
of the sample and the uniaxial-pressure cell were measured by SQUID magnetometer
and the magnetization of the uniaxial-pressure cell, though small ($1-10\%$), were
subtracted from the total magnetization to obtain the precise value of the
sample magnetization.\\
%
\section{Experimental Results}
\begin{figure}[htb]
\begin{center}
\includegraphics[width=8.5cm]{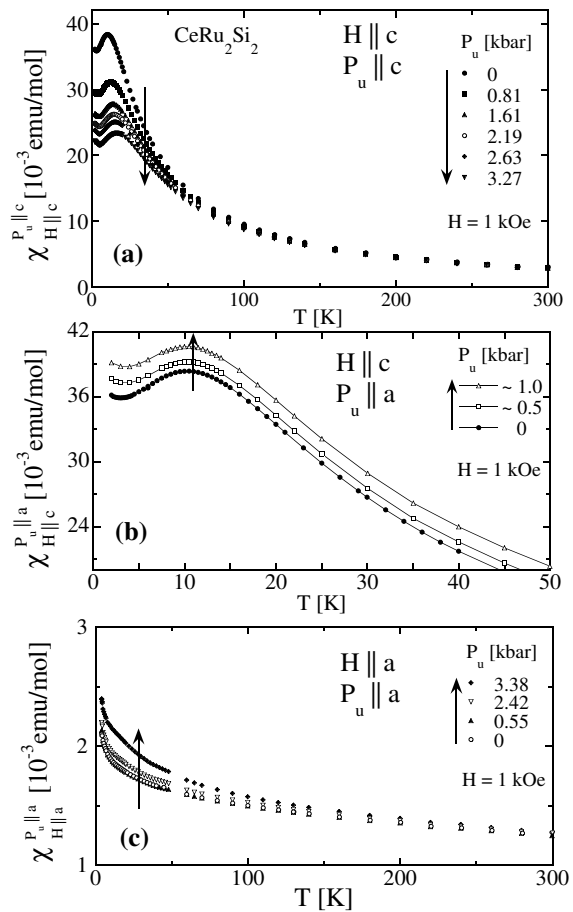} 
\caption{Temperature dependence of magnetic susceptibility ($\chi$) in CeRu$_2$Si$_2$
under uniaxial pressure ($P_u$). (a) $\chi_{H\parallel c}^{P_u\parallel c}(T)$
for $P_u$ applied parallel with the magnetic field ($H$) along the tetragonal $c$ axis.
(b) $\chi_{H\parallel c}^{P_u\parallel a}(T)$ for $P_u$ along the $a$ axis and $H$ along
the $c$ axis. (c) $\chi_{H\parallel a}^{P_u\parallel a}$ for $P_u$
and $H$ along the $a$ axis. The arrows indicate the change of $\chi$
with increasing $P_u$.}
\label{rupparahxt}
\end{center}
\end{figure}
\begin{figure}[htb]
\begin{center}
\includegraphics[width=8.5cm]{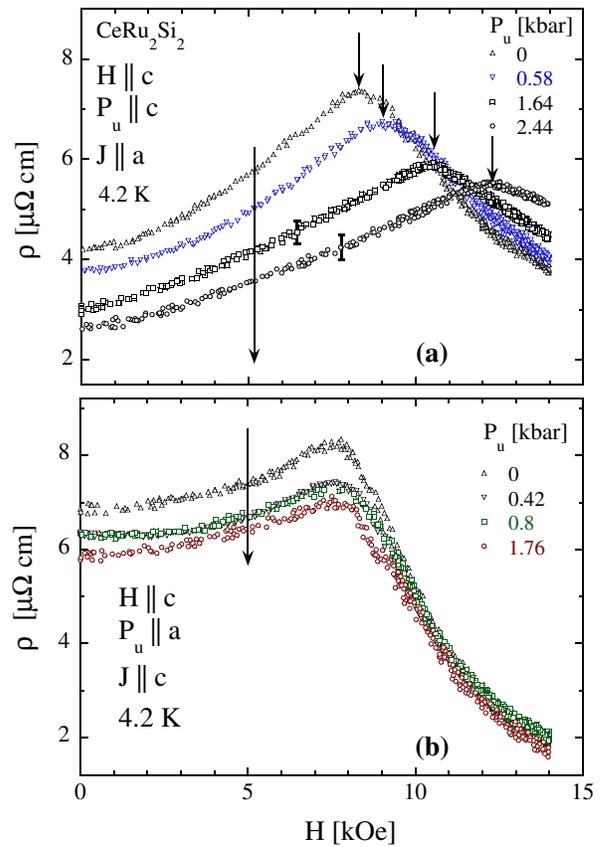}
\caption{Magnetoresistance (MR) in CeRu$_2$Si$_2$ at 4.2 K under uniaxial pressure $P_u$
(a) for $P_u \parallel c$ axis (0 $\leq P_u \leq$ 2.44 kbar) and transverse geometry
($H\parallel c$ axis, $J\parallel a$ axis). The short arrows indicate
the magnetic fields $H_M$ where the maximum in $\rho (H)$ occurs, reflecting the
metamagnetic anomaly. The error bars are put in the vertical axis after smoothing
the scattered data points at higher pressures due to deterioration of the spotweld of the
voltage leads. (b) for $P_u\parallel a$ axis
(0 $\leq P_u \leq$ 1.76 kbar) and longitudinal geometry ($H\parallel J\parallel c$ axis).
The long arrows indicate the decrease of $\rho$ below $H_M$ with increasing
$P_u$ in both figures.}
\label{rutrasmr}
\end{center}
\end{figure}
\begin{figure}[ht]
\begin{center}
\includegraphics[width=8.5cm]{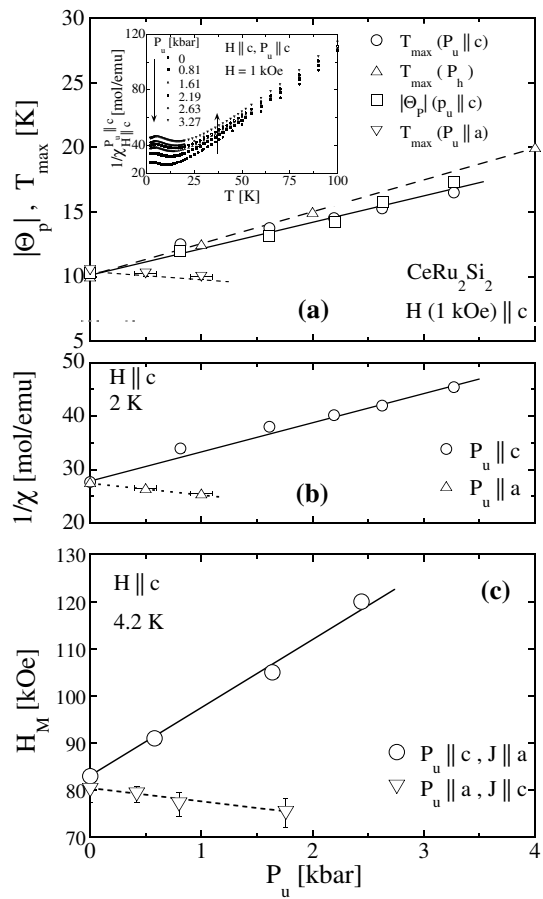}
\caption{(a) Uniaxial pressure dependence of $T_{\rm max}$ in
$\chi_{H\parallel c}^{P_u\parallel c}$ and the absolute value of the
paramagnetic Weiss temperature $\Theta_p$,
estimated from the linear region of the inverse $\chi_{H\parallel c}^{P_u\parallel c}$
plot below 100 K shown in the inset; the evolution of $T_{\rm max}$ under
$P_h$ reproduced from the report by Voiron {\it et al.} (Ref. 7)
is also ploted at the same time. (b) Uniaxial-pressure dependence of the inverse
$\chi_{H\parallel c}^{P_u\parallel c}$ and  $\chi_{H\parallel c}^{P_u\parallel a}$
at 2 K. (c) Uniaxial pressure dependence of $H_M$ in the transverse
($H\parallel c$ axis, $J\parallel a$ axis)
magnetoresistance under $P_u\parallel c$-axis and longitudinal
($H\parallel J\parallel c$ axis) magnetoresistance under $P_u\parallel a$ axis
in CeRu$_2$Si$_2$. The solid and broken lines are guides to eyes.}
\label{rutmaxppara}
\end{center}
\end{figure}
\begin{figure}[ht]
\begin{center}
\includegraphics[width=8.5cm]{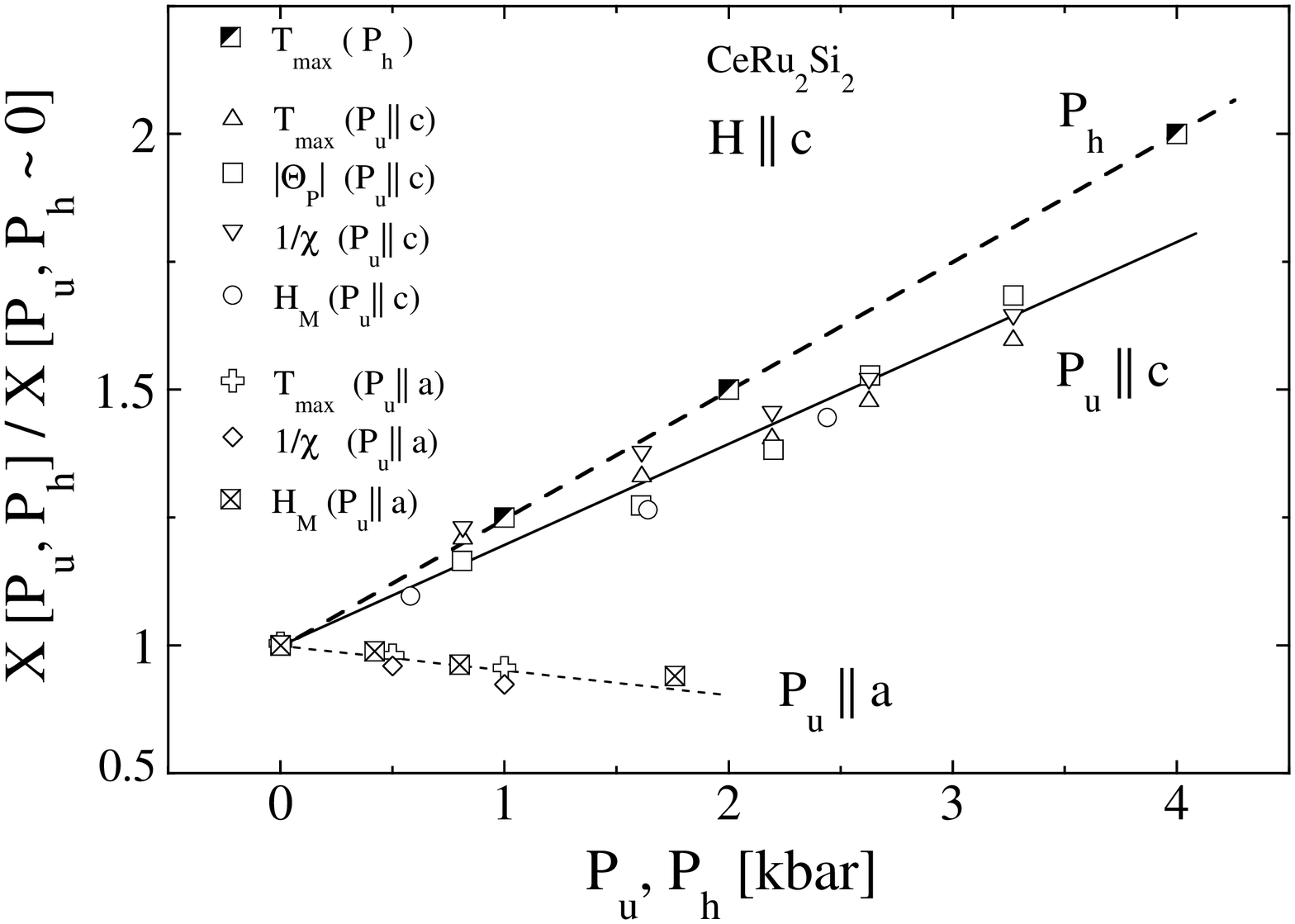}
\caption{$T_{\rm max}$, $\Theta_p$, $\chi_{T \simeq 0}^{-1}$,
and $H_M$ normalized by the ambient pressure value of each parameter as a
function of $P_u$. Solid lines are drawn to show that all the normalized
parameters ($X$) for the same uniaxial-pressure geometry fall roughly on the same line.
$T_{\rm max}$ under $P_h$, reported in Ref. 7 normalized by the ambient pressure value
is also ploted at the same time. The solid and broken lines are guides to the eyes.}
\label{scaling}
\end{center}
\end{figure}

~~~Figure ~\ref{rupparahxt} shows the effect of uniaxial pressure on the temperature
dependence of magnetic susceptibility $\chi (T)$ in CeRu$_2$Si$_2$; 
Fig.~\ref{rupparahxt}(a) is for $P_u \parallel H \parallel c$ axis,
Fig.~\ref{rupparahxt}(b) is
for $P_u\parallel a$ and $H \parallel c$ axis, and Fig.~\ref{rupparahxt}(c) is for
$P_u \parallel H \parallel a$ axis. At ambient pressure, the magnetic susceptibility
is strongly anisotropic depending on whether $H$ is applied parallel or
perpendicular to the tetragonal $c$ axis.
$\chi_{H\parallel c}(T)$ obeys the Curie-Weiss law above $\sim$ 70 K and
then shows a maximum at a temperature $T_{\rm max}$ $\simeq$ 10 K, which is considered
to provide a measure of the Kondo temperature $T_{\rm K}$.
$\chi_{H\parallel a}(T)$ also follows the Curie-Weiss law above 100 K, and
it exhibits much smaller values than $\chi_{H\parallel c}(T)$, without
showing any maximum. The anisotropy ratio
$\chi_{H\parallel c}(T)$/$\chi_{H\parallel a}(T)$
increases largely with decreasing temperatures. These behaviors at ambient pressure
are consistent with those reported before \cite{voiron1}.
When the uniaxial pressure of $P_u\parallel c$ axis is applied,
$\chi_{H\parallel c}^{P_u\parallel c}$ for $H\parallel c$-axis
is largely suppressed at low temperatures with increasing $P_u$, though the effect
of $P_u$ is small at higher temperature ($\geq$100 K).
$T_{\rm max}$ shifts to higher temperatures with increasing $P_u$, i.e.,
$T_{\rm max}\simeq$ 16.5 K at $P_u\simeq$ 3.27 kbar.
The pressure dependence is almost similar to that
under hydrostatic pressure ($P_h$) repoted in Refs. 7 and 8.
By contrast, when $P_u$ is applied perpendicular to the
magnetic field, i.e., $P_u \parallel a$ axis,
$\chi_{H\parallel c}^{P_u\parallel a}$ at
low temperatures is enhanced and $T_{\rm max}$
is slightly suppressed [see Fig.~\ref{rupparahxt}(b)].
On the other hand for $H\parallel a$ axis, $\chi_{H\parallel a}^{P_u\parallel a}$
slightly increases at low temperatures with $P_u\parallel a$ axis
[see Fig.~\ref{rupparahxt}(c)]. This increase of $\chi_{H\parallel a}^{P_u\parallel a}$
is also in contrast to the decrease of $\chi_{H\parallel a}^{P_{h}}$ under
hydrostatic pressure\cite{voiron1}. In the case of $P_u\parallel c$ axis,
$\chi_{H\parallel a}^{P_u\parallel c}$ weakly enhances at low temperatures with
$P_u$ (not shown); it is difficult to separate accurately the background contribution
from the pressure cell which is comparable to the sample magnetization
for $H\parallel a$ axis.\\
\begin{figure}[h]
\begin{center}
\includegraphics[width=8.5cm]{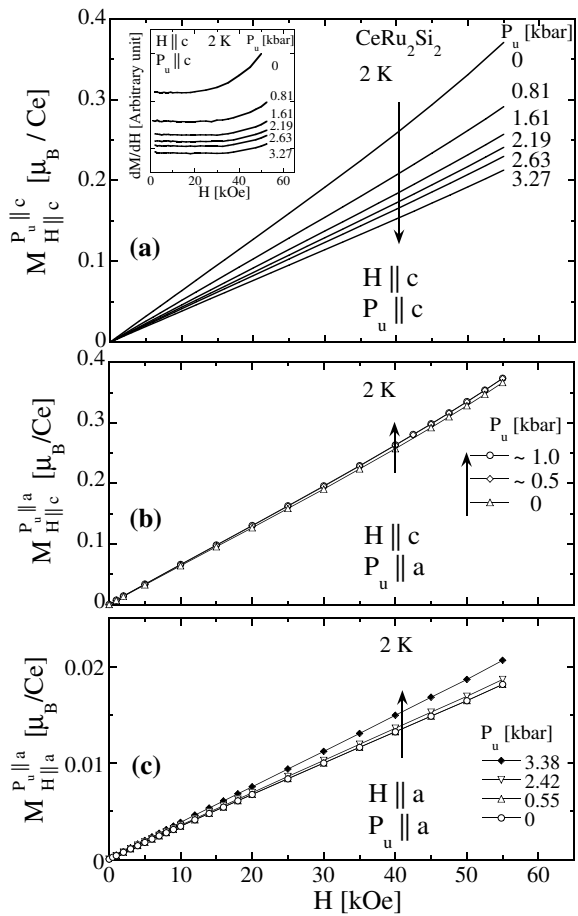}
\caption{Magnetic-field dependence of isothermal magnetization $M (H)$ in CeRu$_2$Si$_2$
under uniaxial pressure at 2 K. (a) $M_{H\parallel c}^{P_u\parallel c}(H)$
for $P_u$ applied parallel with the magnetic field along the tetragonal $c$ axis.
(b) $M_{H\parallel c}^{P_u\parallel a}(H)$ for $P_u$ along the $a$ axis and $H$
along the $c$ axis. (c) $M_{H\parallel a}^{P_u\parallel a}(H)$ for $P_u$ and $H$ along
the $a$ axis. The arrows indicate the change of $M$
with increasing $P_u$. The solid lines are guides to the eyes.}
\label{rumhppara}
\end{center}
\end{figure}
~~~Figure~\ref{rumhppara} shows the effect of uniaxial pressure on the magnetic-field
dependence of isothermal (at 2 K) magnetization $M (H)$ in CeRu$_2$Si$_2$.
The ambient pressure data are consistent with those reported before \cite{voiron1}.
For $H\parallel c$ axis, a positive curvature of $M_{H\parallel c}^{P_u\parallel c}(H)$
is observed as a precursor to the metamagnetic anomaly at $H_M\simeq$ 80 kOe
at ambient pressure. Under $P_u\parallel c$ axis, $M_{H\parallel c}^{P_u\parallel c}$
up to 55 kOe drastically decreases [see Fig.~\ref{rumhppara}(a)] and the curvature
$\partial M_{H\parallel c}^{P_u\parallel c}/\partial H$ at 55 kOe
is also largely suppressed with increasing $P_u$ as shown in the inset
of Fig.~\ref{rumhppara}(a). These facts indicate a shift of the metamagnetic anomaly to
higher magnetic fields by $P_u\parallel c$ axis; actually, this is confirmed by the
magnetoresistance measurement (see Fig.~\ref{rutrasmr}).
This effect of $P_u\parallel c$ axis on the isothermal
$M_{H\parallel c}^{P_u\parallel c}(H)$ is similar
to the effect of $P_h$ on the isothermal magnetization $M_{H\parallel c}^{P_{h}}(H)$
reported in Ref. 7, where $M_{H\parallel c}^{P_{h}}$
and $\partial M_{H\parallel c}^{P_{h}}/\partial H$
also drastically decrease with increasing $P_h$. By contrast, under
$P_u \parallel a$ axis, $M_{H\parallel c}^{P_u\parallel a}$ is enhanced
[see Fig.~\ref{rumhppara}(b)] and
the $\partial M_{H\parallel c}^{P_u\parallel a}/\partial H$ at 55 kOe is also weakly
enhanced. On the other hand for $H\parallel a$ axis, $P_u\parallel a$ axis enhances
$M_{H\parallel a}^{P\parallel a}$ [see Fig.~\ref{rumhppara}(c)]. This increase of
$M_{H\parallel a}^{P_u\parallel a}$ is in sharp contrast to the decrease
of $M_{H\parallel a}^{P_h}$ under hydrostatic pressure\cite{voiron1}.\\
%
~~~In order to see the effect of uniaxial pressure on the MA, we have measured
the magnetoresistance (MR) up to 140 kOe under $P_u$. Figure ~\ref{rutrasmr}(a) shows
the effect of $P_u\parallel c$ axis on the magnetic field dependence of the
transverse magnetoresistance, i.e., $H$ along the 
$c$ axis and the current $J$ along the $a$ axis.
The data for $P_u\sim$ 0 kbar are in close agreement with that reported in Ref. 8.
The resistivity increases with magnetic fields and then shows a maximum or peak at
$H_{\rm max}\simeq$ 83 kOe. The MA manifests itself in this peak at $H_{\rm max}$
($\equiv H_M$). With increasing $P_u$, the resistivity below $H_M$
decreases and the peak shifts to higher $H$ as marked by vertical arrows.
These behaviors are similar to those under hydrostatic
pressure reported in Ref. 8, though there are small quantitative
differences. The most significant differences are that the absolute value of
$\rho$ at the peak (at $H_M$) decreases and the shape of the peak broadens
with increasing $P_u\parallel c$-axis, while these remain almost
unchanged for all values of $P_h$. These behaviors under $P_h$
have been ascribed to the implication that the quasiparticle DOS
always reaches the same critical value at $H_M$\cite{mignot12}.
This conclusion was made by taking into account the following facts:
$H_M$ depends on pressure but it depends only slightly on temperature.
At all values of $P_h$, the $\rho$ at $H = H_M$ of the corresponding $P_h$ shows
almost the same temperature dependence. Therefore, the slope of each $\rho (T)$
curve of $H = H_M$, which gives an estimation
about the DOS at $H = H_M$, remains unchanged with $P_h$ \cite{mignot12}.
Based on the same model, in the present case under $P_u\parallel c$ axis,
the decrease and broadening of the peak at $H_M$ may
indicate a change of the quasiparticle DOS near the metamagnetic
anomaly at $H_M$ with increasing $P_u\parallel c$ axis. However, measurements at
low temperatures are necessary in order to reach a decisive conclusion.\\
~~~Figure ~\ref{rutrasmr}(b) shows the effect of $P_u\parallel a$ axis on the
magnetic field dependence of longitudinal magnetoresistance, i.e., $H$ along $c$ axis
and $J$ along the $a$ axis. The application of $P_u\parallel a$ axis is very
risky, since large cracks easily develop into the samples.
Some samples were completely separated
into pieces in the $c$ plane under pressure, making the measurement impossible
several times. The ambient pressure ($P_u\sim$ 0) data cannot be
compared with that of Ref. 8, since there is no data
for this geometry, however, the data are consistent in
nature to the reported longitudinal MR at lower temperatures
by Kambe {\it et al.} \cite{kambe12}. The data for $P_u\sim$ 0 
and 0.8 kbar have been taken on the same sample piece, but for
$P_u\sim$ 0.4 and 1.76 kbar the data have been taken on two different
sample pieces, although all the pieces were cut from the adjacent part
of the same crystal. The results shown in Fig.~\ref{rutrasmr}(b)
indicate that the sample dependence is minor at least for these three pieces.
With increasing $P_u\parallel a$ axis, the resistivity
decreases for all $H$, though the change in $\rho$ above $H_M$ is smaller than
that below $H_M$. The magnetoresistance peak at $H_M$ ($\sim$ 80 kOe at
ambient pressure), manifesting the metamagnetic anomaly, changes only slightly
with increasing $P_u$. There is a slight decreasing tendency of
$H_M$ [see Fig.~\ref{rutmaxppara}(c)]
with increasing $P_u$, although it is difficult to determine accurately the rate
$\Delta H_M/ \Delta P_u$ due to its extremely small change relative to the
experimental accuracy.\\
~~~Figure~\ref{rutmaxppara}(a) shows the dependence of $T_{\rm max}$ 
in $\chi_{H\parallel c}(T)$ on $P_u$ along with that on $P_h$ reported
by Voiron {\it et al.}\cite{voiron1}. $T_{\rm max}$ is enhanced with
$P_u\parallel c$ axis as ${\rm d}T_{\rm max}/{\rm d}P_u\simeq$ 
1.85 $\pm$ 0.1 K/kbar which is close to the rate 
${\rm d}T_{\rm max}/{\rm d}P_h \simeq$ 2.5 K/kbar for the $P_h$
reported in Refs. 7 and 8. The absolute value of the paramagnetic Weiss temperature
$\Theta_{\rm p}$, determined from the linear region of the inverse of
$\chi_{H\parallel c}^{P_u\parallel c}(T)$ plot (below 100 K) shown in the inset of
Figure~\ref{rutmaxppara}(a), is also found to increase with $P_u\parallel c$ axis
as shown in the same figure. The negative sign of $\Theta_p$ can be attributed
to Kondo correlations. On the other hand, $T_{\rm max}$ is suppressed by
${P_u\parallel a}$ axis. Fig.~\ref{rutmaxppara}(b) shows the $P_u$ dependence of
1/$\chi_{H\parallel c}$ at $T=$ 2 K, which is enhanced and suppressed
by ${P_u\parallel c}$ axis and ${P_u\parallel a}$ axis,
respectively. Note that the low-temperature value of $\chi^{-1}$ is proportional to the
Kondo temperature $T_K$ in the Fermi-liquid regime.
Figure~\ref{rutmaxppara}(c) shows the $P_u$ dependence of $H_M$
in MR. $H_M$ strongly increases by ${P_u\parallel c}$ axis as
$\Delta H_M/\Delta P_u \sim 17 \pm$ 2 kOe/kbar which is close to
the rate $\sim$ 20 kOe/kbar under $P_h$ \cite{mignot12}, while $H_M$
slightly decreases by ${P_u\parallel a}$ axis.\\
~~~Mignot {\it et al.}\cite{mignot12} have shown that the hydrostatic pressure
dependence of $T_{\rm max}$, $\chi_{T \simeq 0}^{-1}$, and $H_M$ show a scaling
behavior. A similar plot has been made as a function of $P_u$, where all the quantities
$T_{\rm max}$, $\Theta_{\rm p}$, $\chi_{T \simeq 0}^{-1}$, and $H_M$ are
normalized by their ambient pressure values as shown in Fig.~\ref{scaling}. $T_{\rm max}$
under $P_h$ normalized by the ambient pressure value is also plotted for comparison.
It is clear that all the quantities scale roughly falling on the same lines. This fact
indicates, similar to the case of hydrostatic pressure\cite{mignot12}, an existence of
a single energy parameter that determines the low temperature properties
controlled by $P_u$. Considering that both the
low-temperature value of $\chi_{T \simeq 0}^{-1}$ in the Fermi-liquid regime and
$\Theta_p$ are proportional to the Kondo temperature $T_K$, the observed
scaling strongly indicates that both $T_{\rm max}$ and $H_M$ are also related
to the Kondo effect. The strong increase (weak decrease) of all the quantities indicates
the strong enhancement (the weak suppression) of hybridization between conduction and
$f$ electrons under $P_u\parallel c$ axis ($P_u\parallel a$ axis).
In other words, $T_K$ is strongly increased by $P_u\parallel c$ axis, while it is
decreased by $P_u\parallel a$ axis.\\
%
%
\section{Discussions}
\begin{figure}[htb]
\begin{center}
\includegraphics[width=8.5cm]{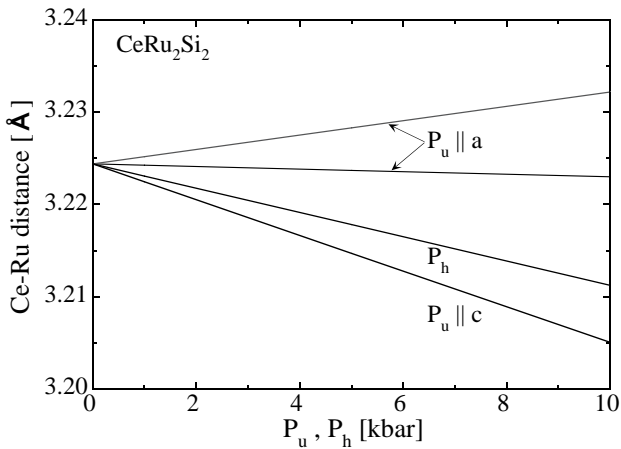}
\caption{Estimated change of Ce-Ru distance in CeRu$_2$Si$_2$ with uniaxial
and hydrostatic pressure using the values of elastic constants (Ref. 14).
The solid lines are guides to the eyes.}
\label{cexdistance}
\end{center}
\end{figure}

~~~In the previous section, we have shown that the effect of uniaxial pressures on the
transport and magnetic properties is quite anisotropic. Especially, a qualitatively
opposite effect on $T_K$ is observed between $P_u\parallel a$ axis and
$P_u\parallel c$ axis configurations. In order to understand this, we estimate the
movement of the surrounding ions relative to the Ce ions due to the uniaxial pressures,
taking into account the elastic properties reported to date.\\
~~~The elastic constants of CeRu$_2$Si$_2$ at 300 K have been reported by
Weber~\cite{weber}, i.e., C$_{11}\simeq$ 2.145, C$_{12}\simeq$ 0.67,
C$_{33}\simeq$ 1.215, and C$_{13}\simeq$ 0.806 Mbar.
We use these values since the low temperature values are not complete; no data are
available at low temperatures for C$_{12}$ in Ref. 14.
Note that the temperature dependence of these quantities does not qualitatively affect
our consequence shown below. Using these values, the linear compressibilities along the
principal crystalline directions are then calculated as
$\kappa_a \equiv - [\Delta a/\Delta P_u]/a \simeq$ 0.63 Mbar$^{-1}$ and
$\kappa_c \equiv - [\Delta c/\Delta P_u]/c \simeq$ 1.33 Mbar$^{-1}$ for the
$a$ axis and $c$ axis, respectively.
Three Poisson ratios are also calculated. Under $P_u\parallel a$ axis, the 
Poisson ratio along the $c$ axis is $\nu_c^{P_u\parallel a}\simeq$ 0.608,
while that along another
$a$ axis is $\nu_a^{P_u\parallel a}\simeq$ 0.084. On the other hand under
$P_u\parallel c$ axis, the Poisson ratio is $\nu_a^{P_u\parallel c}\simeq$ 0.286.
The values are quite anisotropic, i.e., for $P_u$ along the $a$ axis, the
elongation along the $c$ axis is more than seven times larger
than that along the other $a$ axis.\\
~~~It is believed that the hybridization between $f$ states and conduction electrons
in CeRu$_2$Si$_2$ is mainly governed by $d$-$f$ hybridization\cite{endstra}.
The band-structure calculation\cite{yamagami} indicates that there
are five bands crossing the Fermi level (four hole sheets and one electron sheet), and
they consist dominantly of the Ce 4$f$ and the Ru 4$d$ components.
Actually, the nearest atom from Ce ions is Ru, suggesting strong hybridization
between 4$f$ and 4$d$ electrons. Using the same lattice constants and atomic
positions used in Ref.~16, the
distances between the nearest neighbor Ce-Ru are calculated as a function of
uniaxial pressure. The results are shown in Fig.~\ref{cexdistance}.
For the cases of applied uniaxial pressure along
the $c$-axis and hydrostatic pressure, Ce-Ru distance $d_{\rm Ce-Ru}$
(= 3.224 ${\rm \AA}$ at ambient pressure) decreases.
On the other hand for the uniaxial pressure along the $a$ axis, the crystal symmetry
decreases from tetragonal to orthorhombic and two unequivalent $d_{\rm Ce-Ru}$
appear, however, the average value of $d_{\rm Ce-Ru}$ weakly increases.
In the former case, decreasing $d_{\rm Ce-Ru}$ should cause an increase of the $d$-$f$
exchange interaction $J_{df}$, and also an increase of $T_K$, which depends on
$J_{df}$ essentially as $\propto \exp [-1/J_{df} N(E_F)]$, where $N(E_F)$ represents
the density of states on the Fermi energy and is less sensitive to pressure
than $J_{df}$. On the contrary, the uniaxial pressure along the $a$ axis should decrease
$J_{df}$ and hence $T_K$. These expectations are qualitatively consistent
with the present observation. Especially, similar behavior observed for the cases of
$P_u\parallel c$ axis and hydrostatic pressure experiments are naturally explained
as a consequence of the similar pressure dependencies of $d_{\rm Ce-Ru}$.\\
~~~The large anisotropy in the magnetic susceptibility in CeRu$_2$Si$_2$ is
ascribed to the crystalline electric field (CEF) effect. At ambient pressure,
the anisotropy $\chi_{H\parallel c}$/$\chi_{H\parallel a}$ $\sim$ 15 at 4.2 K
suggests the ground state is
$a|\pm\frac{5}{2}\>-\surd\overline{(1-a^2)}|\mp\frac{3}{2}\rangle$,
$a\sim0.96$.\cite{hanzawa1ohara1} The maximum anisotropy in the magnetic susceptibility
is expected for a CEF ground state of pure $|\pm\frac{5}{2}\rangle$.
$\chi_{H\parallel c}$ is largely suppressed, while
$\chi_{H\parallel a}$ is slightly enhanced by $P_u\parallel c$ axis.
Therefore, the anisotropy $\chi_{H\parallel c}/\chi_{H\parallel a}$
decreases with increasing $P_u\parallel c$ axis. This may reflect a
change of the CEF ground state, i.e., an increase of the $|\mp\frac{3}{2}\rangle$
component in the ground state by $P_u\parallel c$ axis. For the Ce$^{3+}$ ion,
the charge distribution of
$|\mp\frac{3}{2}\rangle$ is dumbbell shaped with its elongation along the
$c$-axis \cite{walter}.
Moreover in the CeRu$_2$Si$_2$ crystal unit cell, the Ru atom is located  at
($0, \frac{1}{2}, \pm\frac{1}{4}$) with respect to the Ce ion.\cite{yamagami}
Therefore, besides the decrease of Ce-Ru distance,
an increase of the $|\mp\frac{3}{2}\rangle$
component also consistently favors the strong enhancement of Ce-Ru hybridization by
$P_u\parallel c$ axis. It is argued that there is an anomalous peak structure
in the partial density of the hybridized-band state (the DOS) due to anisotropic
hybridization between 4$f$ and conduction electrons in the
case where the lowest CEF level is $|\pm\frac{5}{2}\rangle$.\cite{aoki1,hanzawa1ohara1}
In this case, the anisotropic hybridization has an angular dependence
characterized by (1$-$\^k$^2$$_z$)$^2$. The differential susceptibility
diverges when the peak in the DOS crosses the Fermi level giving rise to the
metamagnetic anomaly at $H_M$. As discussed in the previous section comparing the
transverse MR under $P_u\parallel c$ axis with the reported MR behavior under hydrostatic
pressure, the decrease and broadening of the $\rho$ peak at $H_{\rm max}$ may suggest
a change of DOS with $P_u\parallel c$ axis. A possible change of the CEF ground state,
i.e., an increase of the $|\mp\frac{3}{2}\rangle$ component by $P_u\parallel c$ axis may
cause a decrease and broadening of the anomalous peak in the DOS near the Fermi
level predicted for anisotropic hybridization. It may be noted that
for a pure $|\mp\frac{3}{2}\rangle$ ground state the DOS near the Fermi level
has a finite value rather than a peak as shown in the case of the Kondo insulator
CeNiSn\cite{ikeda1,hanzawa1ohara1}. In order to determine the quantitative change
or broadening of the DOS, measurements of the field dependence of specific heat
under uniaxial pressure are needed.\\
%
\section{Conclusion}
~~~We have found that the uniaxial pressure has an anisotropic
effect on the magnetic and transport properties in the heavy-fermion
metamagnet CeRu$_2$Si$_2$ with a direct influence on the hybridization.
The characteristic parameters $T_{\rm max}$, $\Theta_p$, $\chi_{T \simeq 0}^{-1}$,
and $H_M$ roughly scale as the uniaxial pressure is varied, leading to a
single-energy-scale picture, namely, the variation of $T_K$.
The results suggest that $T_K$ (or $d$-$f$ hybridization) is strongly enhanced
for the pressure along the $c$ axis due to the decrease of the nearest Ce-Ru distance,
while $T_K$ is weakly suppressed for the pressure along the $a$ axis due to
the increase of Ce-Ru distance. The decrease of the anisotropy of
magnetic susceptibility and the decrease and broadening of the magnetoresistance peak
at the metamagnetic anomaly indicate a controlling of the anisotropic hybridization
by uniaxial pressure.\\
%
\begin{center}
{ACKNOWLEDGMENT}\\
\end{center}
~~~The authors are grateful to thank Profs. E. V. Sampathkumaran,
M. Kohgi, O. Sakai, M. Yoshizawa, Y. Uwatoko, and K. Miyake for their comments
and help. This work has been partly supported by a Grant-in-Aid
for Scientific Research from the Ministry of Education, Science, Sports, and
Culture of Japan.
%



\end{document}